 \documentstyle[preprint,prl,tighten,aps,epsfig]{revtex}  %

\begin{document}

\draft

\preprint{\begin{tabular}{l}
\hbox to\hsize{\mbox{ }\hfill KIAS--P99080}\\
\hbox to\hsize{\mbox{ }\hfill hep--ph/9909315}\\
\hbox to\hsize{\mbox{ }\hfill September, 1999}\\
          \end{tabular} }

\title{$s$-Channel Production of Minimal Supersymmetric Standard Model Higgs 
       Bosons at a Muon Collider with Explicit CP Violation}

\author{S.Y. Choi and Jae Sik Lee}
\address{Korea Institute for Advanced Study, 207--43, Cheongryangri--dong
         Dongdaemun--gu, Seoul 130--012, Korea}

\maketitle

\begin{abstract}
A muon collider with controllable energy resolution and transverse
beam polarization provides a powerful probe of the Higgs sector in the 
minimal supersymmetric standard model with explicit CP violation, 
through $s$-channel production of Higgs bosons. The production rates and
the CP--even and CP--odd transverse--polarization asymmetries
are complementary in diagnosing CP violation in the Higgs sector.
\end{abstract}



The experimental observation of Higgs bosons and the
detailed investigation of their fundamental properties are crucial 
for our understanding of the mechanism responsible
for the electroweak symmetry breaking. This crucial quest in particle
physics constitutes the primary reason for having a
$\mu^+\mu^-$ collider (MC) \cite{MUCOL}, the physics potential of
which has been investigated with a considerable amount of effort.  The
fact that muons are significantly heavier than electrons makes a MC
very attractive for both practical and theoretical reasons \cite{S_H}:
(i) synchrotron radiation does not limit their circular acceleration
and multi-TeV energies for a MC can be realized; (ii) the muon beam
energy distribution is not smeared out by beamstrahlung; (iii) the
larger Yukawa couplings of muons in many cases allow for copious
production of Higgs bosons as $s$-channel resonances, making precision
studies of the Higgs sector possible. In particular, one can search
for CP violation in the couplings of Higgs bosons to muons by
measuring the production rates and the polarization asymmetries
constructed with transversely polarized muon beams \cite{SONI}.  In
this paper we point out the possibility of studying loop--induced
CP--violating scalar--pseudoscalar mixing \cite{PW} through
$s$-channel production of neutral Higgs bosons at a MC in the minimal
supersymmetric standard model (MSSM). The experimental
tools assumed to be available for this investigation are controllable
beam energy resolution and beam polarization, for both muons and
anti--muons.

In general, the MSSM has several new CP--violating phases which are
absent in the standard model (SM). Furthermore, it has recently been
pointed out \cite{IN,Pok} that these phases do not have to be suppressed
in order to satisfy the constraints from electron and neutron electric
dipole moments. Of particular interst in this context is the so--called 
effective supersymmetry (SUSY) model \cite{KAPLAN} in which 
sfermions of the first and second generations are decoupled\footnote{
The first and second generation sfermions do not have to be completely 
decoupled. However, only if they are fairly heavy, ${\cal O}($1 TeV), 
the phases can be large without substantial fine--tuning \cite{Pok}.},
but sfermions of the third generation remain relatively light to preserve
naturalness. Motivated
by scenarios of this type, we concentrate on CP violation in the Higgs
sector induced at one--loop level by CP--violating phases in the stop
and sbottom sectors. Characteristic CP--violating phenomena in this
scheme are a possibly large mixing between the CP--even and CP--odd
neutral Higgs bosons and an induced relative phase $\xi$
\cite{PW,DEMIR} between the vacuum expectation values of the two Higgs
doublets.

The dominant source of one--loop radiative corrections to the Higgs
potential of the MSSM are the stop and sbottom sectors due to the large 
$t$ and (for large values of $v_2/v_1\equiv \tan\beta$) $b$ Yukawa couplings. 
Then, the characteristic size of CP 
violation induced in the Higgs sector is determined by the factor
$(1/32\pi^2)(|\mu||A_f|Y_f^4/M^2_{\rm SUSY}) \sin\phi_{\rm CP}$, where
$Y_f$ is the Yukawa coupling of the fermion $f$, $\phi_{\rm CP}={\rm
Arg}(\mu A_f)+\xi$ for $f=t,b$, and $M_{\rm SUSY}$ is a typical
SUSY--breaking scale, the square of which might be taken to be 
the average of two sfermion ($\tilde{f}_1, \tilde{f}_2$) masses 
squared, i.e. $M^2_{\rm SUSY}=(1/2)[m^2_{\tilde{f}_1}+m^2_{\tilde{f}_2}]$. 
In the present work, we take a common phase $\Phi$ for $A_{t,b}$ and present
results for the following parameter values:
\begin{eqnarray}
\label{eq:para}
&& |A_t|=|A_b| = 1~{\rm TeV}\,, \qquad |\mu|=2~{\rm TeV}\,,\nonumber \\
&& M_{\rm SUSY}= 0.5~{\rm TeV}\,, \qquad {\rm Arg}(\mu)+\xi=0 \,.
\label{eq:parameter}
\end{eqnarray}
In addition, noting that the $CP$--violating phases could weaken the present 
experimental bounds on the lightest Higgs mass up to about 60 GeV \cite{PW}, 
we simply apply the lower mass limit $M_{H_1}\geq 70$ GeV
to the lightest Higgs--boson\footnote{For this
bound, the charged Higgs--boson mass is found to be well beyond the 
present experimental lower bound irrespective of the CP--violating phase.}
determining the lowest allowed value of $M_{H^\pm}$ for a given 
set of SUSY parameters.  For the parameter $\tan\beta$, we take $\tan\beta=3$ 
or $30$ as the values representing the small and large $\tan\beta$ cases, 
respectively.
This representative choice (\ref{eq:para}) of parameters, (which was
considered in Ref.~\cite{PW} as well), enables us to avoid the
possible Barr--Zee type EDM constraints \cite{CKP} on the phase $\Phi$.
The Higgs boson masses and couplings to the SM particles are then completely
determined by three independent parameters
$\{\tan\beta,\Phi,M_{H^\pm}\}$.

In the MSSM with explicit CP violation, the couplings of the neutral
Higgs bosons to the SM fermions are significantly affected by
scalar--pseudoscalar mixing \cite{PW}. Specifically, the interactions
of the neutral Higgs bosons with muons are described by the Lagrangian:
\begin{eqnarray}
{\cal L}_{H\mu\mu}=-\frac{em_\mu}{2m_Ws_Wc_\beta}\, 
    \bar{\mu}\left[O_{2,4-i}-i s_\beta O_{1,4-i}\gamma_5\right]\mu\, H_i\,,
\end{eqnarray}
where the transpose of $O$ is the $3\times 3$ orthogonal matrix rotating 
three neutral Higgs bosons $H_\alpha=(a,\phi_d,\phi_u)$ into 
three mass eigenstates $H_i$ ($i=1,2,3$) with the mass ordering
$M_{H_3}\ge M_{H_2} \ge M_{H_1}$: $H_\alpha=O_{\alpha,4-i} H_i$.  The
interaction Lagrangian enables one to derive the helicity amplitudes
${\cal M}_{\lambda\bar{\lambda}}$ for the $s$-channel production of
the neutral Higgs bosons in $\mu^+\mu^-$ collisions, which are given
by
\begin{eqnarray}
{\cal M}^i_{\lambda\bar{\lambda}} = -\frac{e m_\mu M_{H_i}}{2 m_W s_W c_\beta}
          \left(\lambda\beta_i O_{2,4-i}+i s_\beta O_{1,4-i}\right) 
          \delta_{\lambda\bar{\lambda}}\,,
\end{eqnarray}
where $\lambda,\bar{\lambda}=\pm$ and $\beta_i=\sqrt{1-4m_\mu^2/M_{H_i}^2}$.
Then, the unpolarized cross section $\sigma^0_{H_i}$ for the $s$-channel
process $\mu^+\mu^-\rightarrow H_i$ is expressed in terms
of the decay width $\Gamma(H_i\rightarrow\mu\mu)$ as 
\begin{eqnarray}
\sigma^0_{H_i}(\sqrt{s})=\frac{4\pi\Gamma(H_i\rightarrow\mu\mu)\Gamma_{H_i}}{
                               (s-M_{H_i}^2)^2 + M_{H_i}^2 \Gamma_{H_i}^2}\,,
\end{eqnarray}
and the polarized $s$-channel cross section is given in a simple form as
\begin{eqnarray}
\sigma_{H_i}=\sigma^0_{H_i}\bigg[1+P_L{\overline P_L} 
            +(P_T{\overline P_T}-P_N{\overline P}_N){\cal Y}_i
            +(P_T{\overline P_N}+P_N{\overline P_T})
                      {\overline{{\cal Y}_i}}\bigg] \,,
\end{eqnarray}
where $P_{L,T,N}$ (${\overline P}_{L,T,N}$) denote the longitudinal,
transverse and normal polarizations of the $\mu^-$ ($\mu^+$) beam
and the transverse--polarization (TP) asymmetries ${\cal Y}_i$ and
${\overline {\cal Y}_i}$ are defined in terms of the helicity amplitudes 
${\cal M}_{\lambda\bar{\lambda}}$ as  
\begin{eqnarray}
{\cal Y}_i&=&\frac{2{\cal R}e({\cal M}^i_{--}{\cal M}^{i*}_{++})}{
             \left|{\cal M}^i_{--}\right|^2+\left|{\cal M}^i_{++}\right|^2} \,,
           \nonumber \\
{\overline {\cal Y}_i}&=&\frac{2{\cal I}m({\cal M}^i_{--}{\cal M}^{i*}_{++})}{
             \left|{\cal M}^i_{--}\right|^2+\left|{\cal M}^i_{++}\right|^2} \,.
\end{eqnarray}
Here, the ${\cal Y}_i$ are CP--even, but the ${\overline {\cal Y}_i}$
are CP--odd. In the CP--invariant theories, they should satisfy the relations;
${\cal Y}_i=\mp 1$ depending on whether $H_i$ is a pure CP--even
or CP--odd state, and ${\overline {\cal Y}_i}= 0$. In other words,
$|{\cal Y}_i|< 1$ and ${\overline {\cal Y}_i}\neq 0$ imply CP violation
directly.

The physical quantities determining the unpolarized cross sections are
the total widths of the Higgs bosons, their partial decay widths
$\Gamma(H_i\rightarrow \mu\mu)$, as well as their masses. The Higgs
masses, in particular, the lightest Higgs boson mass are expected to
be measured with an error of less than 100 MeV at $e^+e^-$ or
$\mu^+\mu^-$ colliders \cite{HMASS}. So, it is possible to probe Higgs
boson physics by setting the beam energy very close to the Higgs
mass. In this case, $s$-channel studies of narrow Higgs resonances
depend critically on the beam energy resolution $R=\Delta E_{\rm
beam}/E_{\rm beam}$ with respect to the resonance width
$\Gamma_{H_i}$. The energy spectrum of each beam is expected to be
Gaussian to a good approximation with an rms deviation $\sigma_E \sim
2~{\rm MeV} \left(\frac{\sqrt{s}}{100~{\rm GeV}}\right)
\left(\frac{R}{0.003~\%}\right)$ and with a capability of changing $R$
from $0.01\%$ to $1\%$. Of crucial importance are then the ratios of
the rms error $\sigma_E$ for the given $R$ and the calculated total
widths of Higgs bosons. According to a detailed analysis of Higgs
boson decays \cite{SYCHOI}, the widths of the heavier Higgs states
$H_{2,3}$ are much larger than the width\footnote{For a large value of 
$\tan\beta$, $\Gamma_{H_1}$ can be as large as 1 GeV which requires a 
larger $R$ to satisfy $\sigma_E\gg \Gamma_{H_1}$.}   
of the lightest
Higgs boson $H_1$, which is of the order of MeV for a small value of
$\tan\beta$. It will thus be reasonable to take an energy resolution
$R$ such that $\sigma_E\gg\Gamma_{H_1}$ and
$\sigma_E\ll\Gamma_{H_{2,3}}$.

The effective cross section $\overline{\sigma}_{H_i}$ is given by
a convolution with the Gaussian beam energy distribution: 
\begin{eqnarray}
\overline{\sigma}_{H_i}=\int\sigma_{H_i}\left(\sqrt{s'}\right)
                        \frac{\exp[-(\sqrt{s'}-\sqrt{s})^2/2\sigma_E^2]}{
                              \sqrt{2\pi}\sigma_E} ~d\sqrt{s'}\,.
\end{eqnarray}
For $\sigma_E$ satisfying $\sigma_E\gg\Gamma_{H_1}$ and 
$\sigma_E\ll\Gamma_{H_{2,3}}$, the effective unpolarized cross sections 
$\overline{\sigma}^0_{H_i}$ at $\sqrt{s}=M_{H_i}$ $(i=1,2,3)$ are given
by
\begin{eqnarray}
&& \overline{\sigma}^0_{H_1}= \frac{\pi\Gamma_{H_1}}{2\sqrt{2\pi}\sigma_E}
                          ~\sigma^0_{H_1}(M_{H_1}) 
                        =  \frac{2\pi^2}{\sqrt{2\pi}\sigma_E}
                           \frac{\Gamma(H_1\rightarrow\mu\mu)}{M^2_{H1}}\,,
                           \nonumber \\
&& \overline{\sigma}^0_{H_{2,3}}= \sigma^0_{H_{2,3}}(M_{H_{2,3}})
                        =  \frac{4\pi}{M^2_{H_{2,3}}}
                           {\cal B}(H_{2,3}\rightarrow \mu\mu)\,, 
\end{eqnarray}
to a good approximation. Therefore, measurements of the effective 
unpolarized cross sections yield $\Gamma(H_1\rightarrow\mu\mu)$ and 
${\cal B}(H_{2,3}\rightarrow\mu\mu)$. In addition, 
it is possible to measure the polarization--dependent quantities 
${\cal Y}_i$ and $\overline{{\cal Y}_i}$ by use of transversely--polarized 
muon and anti--muon beams, giving independent information on the scalar 
and pseudoscalar couplings of the Higgs bosons to muons.

As shown explicitly in Ref.~\cite{SYCHOI}, ${\cal
B}(H_1\rightarrow\mu\mu)$ is almost independent of $\Phi$, but
$\Gamma(H_1\rightarrow\mu\mu)$ (as well as $\Gamma_{H_1}$) is strongly
dependent on $\Phi$. 
On the contrary, ${\cal B}(H_{2,3}\rightarrow\mu\mu)$ depend significantly 
on $\Phi$, but $\Gamma(H_{2,3}\rightarrow\mu\mu)$ are almost 
independent of $\Phi$. The effective unpolarized cross
sections are thus expected to be very sensitive to $\Phi$ because their
measurements correspond to those of the strongly phase--dependent
quantities $\Gamma(H_1\rightarrow\mu\mu)$ and
${\cal B}(H_{2,3}\rightarrow\mu\mu)$.  In the following, we will
demonstrate these encouraging aspects in detail.

{\it The lightest Higgs boson}$-$ In our numerical analysis, we take
$R=0.15\%$, which corresponds to $\sigma_E\sim 0.1~{\rm GeV}$ for
$M_{H_1}=100~{\rm GeV}$. As shown in Fig.~1, for a fixed CP--violating
phase the effective unpolarized
cross section $\overline{\sigma}^0_{H_1}$ decreases rapidly with
increasing $M_{H_1}$, approaching the (phase--independent) cross
section for the production of the SM Higgs boson. However,
for most of the lightest Higgs--boson mass $M_{H_1}$, the cross section 
$\overline{\sigma}^0_{H_1}$ depends strongly 
on the CP--odd phase $\Phi$, i.e. on the charged Higgs--boson
mass, except for small $M_{H_1}$ for $\tan\beta=30$. 
However, as far as the unpolarized cross section is concerned, changing
the soft stop/sbottom sector parameters could mimic the effects
of changing the CP--violating phase. Thus, it is crucial to 
confirm the existence of the finite CP--violating phase {\it directly}
through the TP asymmetries ${\cal Y}_1$ and $\overline
{\cal Y}_1$. Fig.~2 shows the CP--even and CP--odd asymmetries  versus 
$M_{H_1}$, the size of which is independent of $R$.
These TP asymmetries are generally very sensitive to the phase $\Phi$
for both $\tan\beta=3$ and $\tan\beta=30$. The only exception again
occurs in the ``decoupling limit'' $M_{H^+} \rightarrow \infty$, where
$M_{H_1}$ approaches its upper limit and the TP asymmetries take their
SM values (${\cal Y}_{SM} = -1, \ \overline{\cal Y}_{SM} =
0$). Note also that, except in the decoupling limit, the total cross
section scales $\propto \tan^2 \beta$, while the $\tan \beta$
dependence of the TP asymmetries is relatively mild.  In this light,
the unpolarized cross section and the polarization asymmetries play a
complementary role in extracting the values of MSSM parameters.

{\it The heavier Higgs bosons}$-$ The branching ratios ${\cal
B}(H_{2,3} \rightarrow \mu\mu)$ determined independently of $\sigma_E$
($\ll \Gamma_{H_{2,3}}$) are in general strongly dependent on the
phase $\Phi$ for a small value of $\tan\beta$, unlike ${\cal
B}(H_1\rightarrow \mu\mu)$, which remains almost constant. However,
they become (almost) independent of $\Phi$ for large $\tan\beta$ where
fermionic decay modes are by far the dominant ones.  We therefore
study the effective unpolarized cross sections
$\overline{\sigma}^0_{H_{2,3}}$ only for $\tan\beta=3$.  Results are
presented in Fig.~3, which shows $\overline{\sigma}^0_{H_{2,3}}$
versus $M_{H_{2,3}}$, respectively, for various values of $\Phi$.  The
cross sections without CP violation decrease abruptly with increasing
$M_{H_{2,3}}$, especially when new decay channels ($Z H_1, H_1 H_1,
t \bar{t}$) open up, but the monotonic decrease is not retained for
non--vanishing values of the phase $\Phi$. In particular, for small
Higgs boson masses the cross sections are significantly affected by
the phase $\Phi$. For large Higgs boson masses they are not so
strongly dependent on the phase, since here the total widths of the unmixed
heavy scalar and pseudoscalar Higgs bosons are both dominated by the
partial widths into $t \bar{t}$, which become identical for both modes
in the limit of large Higgs boson masses.

Our analysis for the unpolarized cross sections for heavy Higgs bosons
clearly shows the need to use other independent observables to
determine the effects of the CP phase for large values of Higgs
masses.  We therefore investigate the dependence of the CP--even and
CP--odd TP asymmetries ${\cal Y}_{2,3}$ and $\overline{{\cal
Y}_{2,3}}$ on the CP phase. The TP asymmetries ${\cal Y}_2$ and
$\overline{{\cal Y}_2}$ versus $M_{H_2}$ are presented in Fig.~4 and
the TP asymmetries ${\cal Y}_3$ and $\overline{{\cal Y}_3}$ versus
$M_{H_3}$ in Fig.~5, for the parameter set (\ref{eq:para}). We find
three interesting aspects concerning those TP asymmetries: 
(i) for both $\tan\beta=3$ and $\tan\beta=30$ all the TP asymmetries
are very sensitive to the CP phase. This strong dependence will be
essential in determining the CP phase for a large value of $\tan\beta$
and large $M_{H^+}$, because in that case all unpolarized cross
sections (for given Higgs masses) are nearly independent of $\Phi$, as
mentioned before;
(ii) if the CP--even TP asymmetries are small in size, then the CP--odd
TP asymmetries are large in size and vice versa. This property stems from
the self--evident sum rule ${\cal Y}^2_i+\overline{\cal Y}^2_i =
1$ for every $i$;
(iii) except for the low mass regime where mixing between the lightest
Higgs state and the heavier Higgs states is enhanced, the CP--even TP
asymmetries of the two heavy states are opposite in sign and so are
the CP--odd TP asymmetries. This reflects the fact that the lightest
Higgs state is (almost) decoupled from the two heavy states, so that
the latter undergo typical two--state mixing.  As the TP asymmetries
have opposite signs for two Higgs bosons, there will be sizable
cancellations if two states are degenerate. Therefore, 
using the TP asymmetries as a powerful probe of the CP phase
requires that the mass difference be at least comparable 
to or larger than their decay widths. Numerically, we have checked 
that even for $\tan\beta=30$ the mass difference is indeed comparable 
to the decay widths in the CP--invariant case and it becomes larger 
for non--trivial values of the CP phase unless the masses are larger
than 500 GeV \cite{PW,SYCHOI}. 
			       

In summary, CP--violating phases in the stop and sbottom sectors
modify the mass spectrum and couplings of the Higgs bosons
significantly from those in the CP--invariant theory.  We have shown
that a controllable energy resolution and beam polarization at a MC
offer powerful and independent opportunities for probing the MSSM
Higgs sector via $s$--channel resonance production even with
loop--induced explicit CP violation. For the very narrow lightest
Higgs--boson resonance, a relatively large beam energy resolution
allows one to measure the muonic decay width precisely, and the
measured width is complementary to the CP--even and CP--odd TP
asymmetries in determining the free parameters.  
However, for a large value of $\tan\beta$ and large $M_{H^+}$, it
turned out that measurements of the transverse polarization
asymmetries of the {\em heavier} Higgs bosons are essential for
determining the CP phase. The availability of transversely polarized
beams at a Muon Collider would therefore be of crucial importance for
precisely probing the characteristics of the Higgs sector in the
general CP--noninvariant theories.

\vskip 0.3cm

\section*{Acknowledgments}

The authors thank Manuel Drees for valuable comments.
The work of SYC was supported by the Korea Science and Engineering 
Foundation (KOSEF) through the KOSEF--DFG large collaboration project, 
Project No.~96--0702--01-01-2.

\begin{figure}
 \begin{center}
 \hbox to\textwidth{\hss\epsfig{file=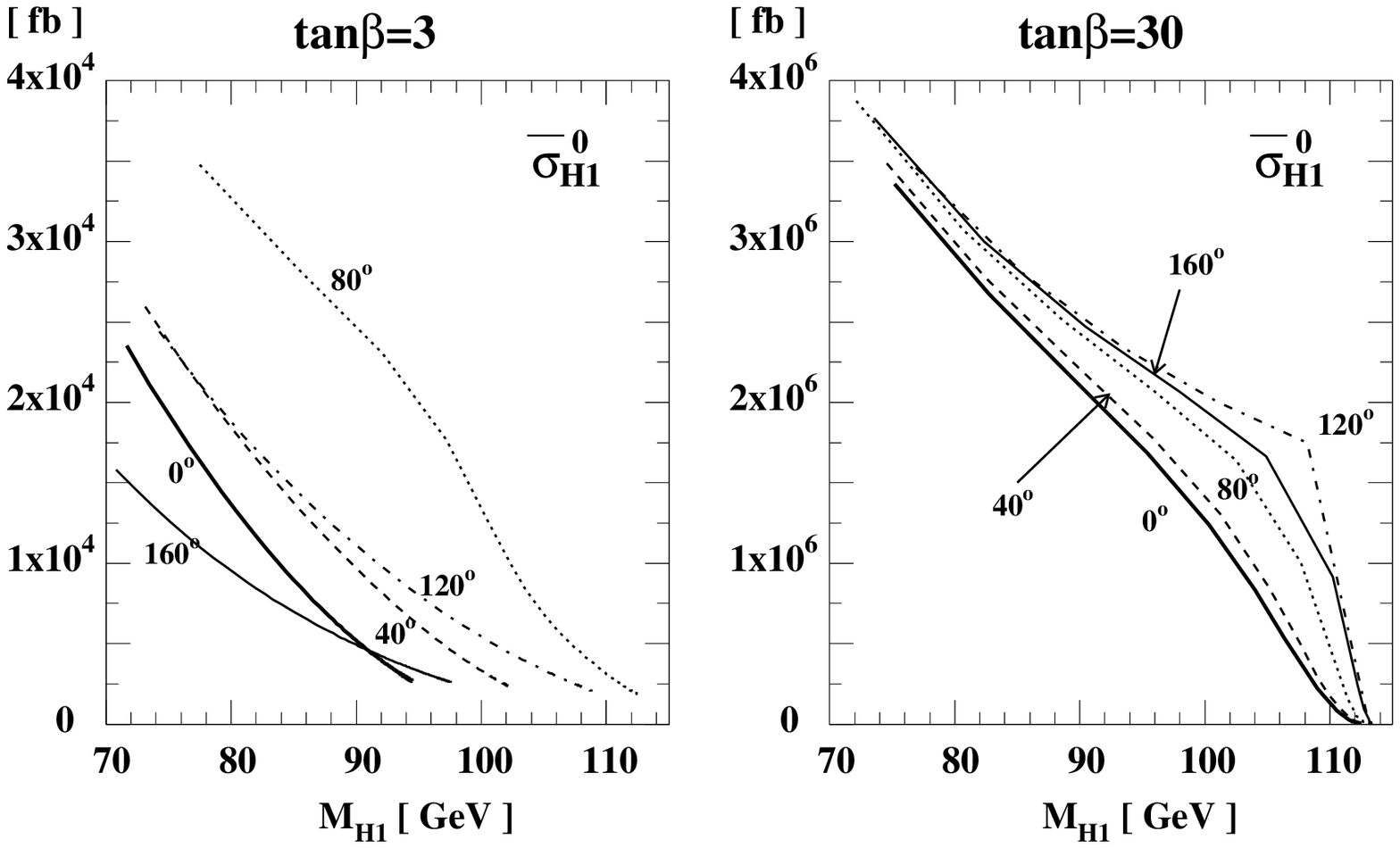,width=16cm,height=14cm}\hss}
 \end{center}
\caption{The effective unpolarized cross section $\overline{\sigma}^0_{H_1}$ 
         versus $M_{H_1}$ with $\Phi=0^0, 40^0, 80^0, 120^0$, and $160^0$
         for $\tan\beta=3$ (left) and $\tan\beta=30$ (right). 
	 Here, $R$ is taken to be $0.15\%$.}
\label{fig1}
\end{figure}
\begin{figure}
 \begin{center}
 \hbox to\textwidth{\hss\epsfig{file=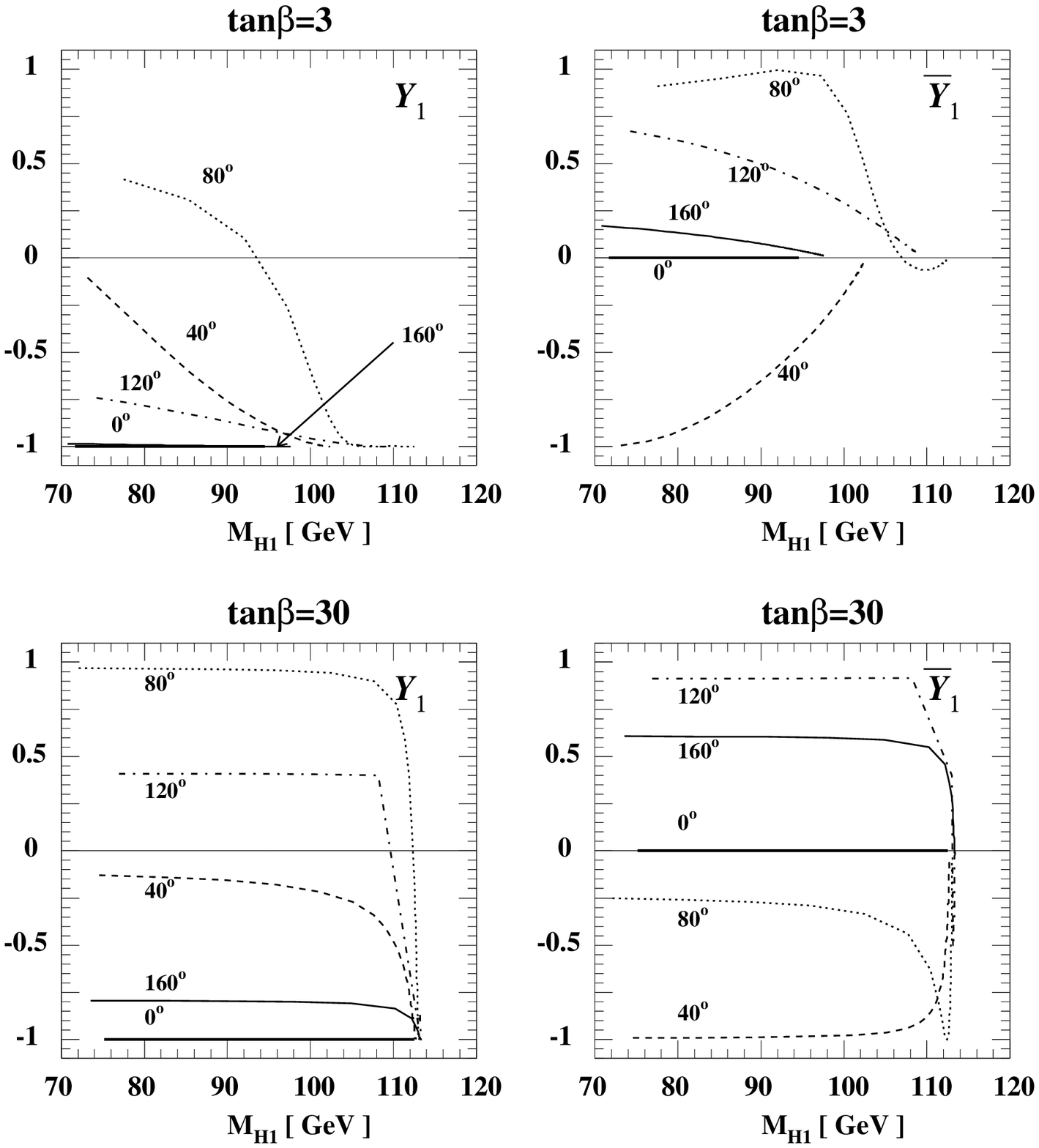,width=16cm,height=14cm}\hss}
 \end{center}
\caption{The CP--even and CP--odd TP asymmetries ${\cal Y}_1$ and
          $\overline{{\cal Y}_1}$ versus $M_{H_1}$ with $\Phi=0^0, 40^0, 80^0, 
          120^0$, and $160^0$ for $\tan\beta=3$ (upper) and $\tan\beta=30$
	  (lower).} 
\label{fig2}
\end{figure}
\begin{figure}
 \begin{center}
 \hbox to\textwidth{\hss\epsfig{file=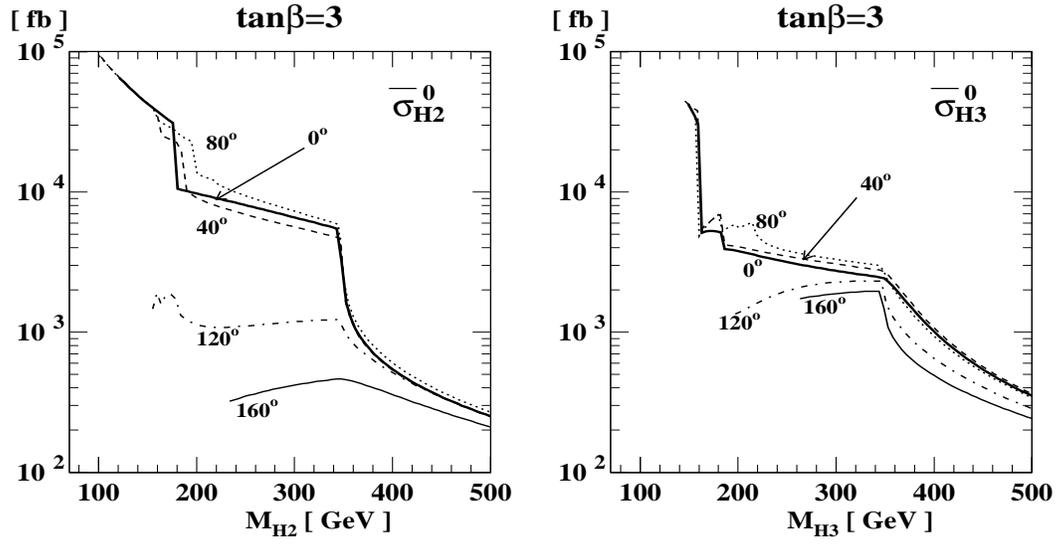,width=16cm,height=14cm}\hss}
 \end{center}
\caption{The effective unpolarized cross sections $\overline{\sigma}^0_{H_2}$ 
         versus  $M_{H_{2}}$ (left) and  $\overline{\sigma}^0_{H_3}$ versus  
         $M_{H_{3}}$ (right) with $\Phi=0^0, 40^0, 80^0, 120^0$, and $160^0$
         for $\tan\beta=3$. }
\label{fig3}
\end{figure}
\begin{figure}
 \begin{center}
 \hbox to\textwidth{\hss\epsfig{file=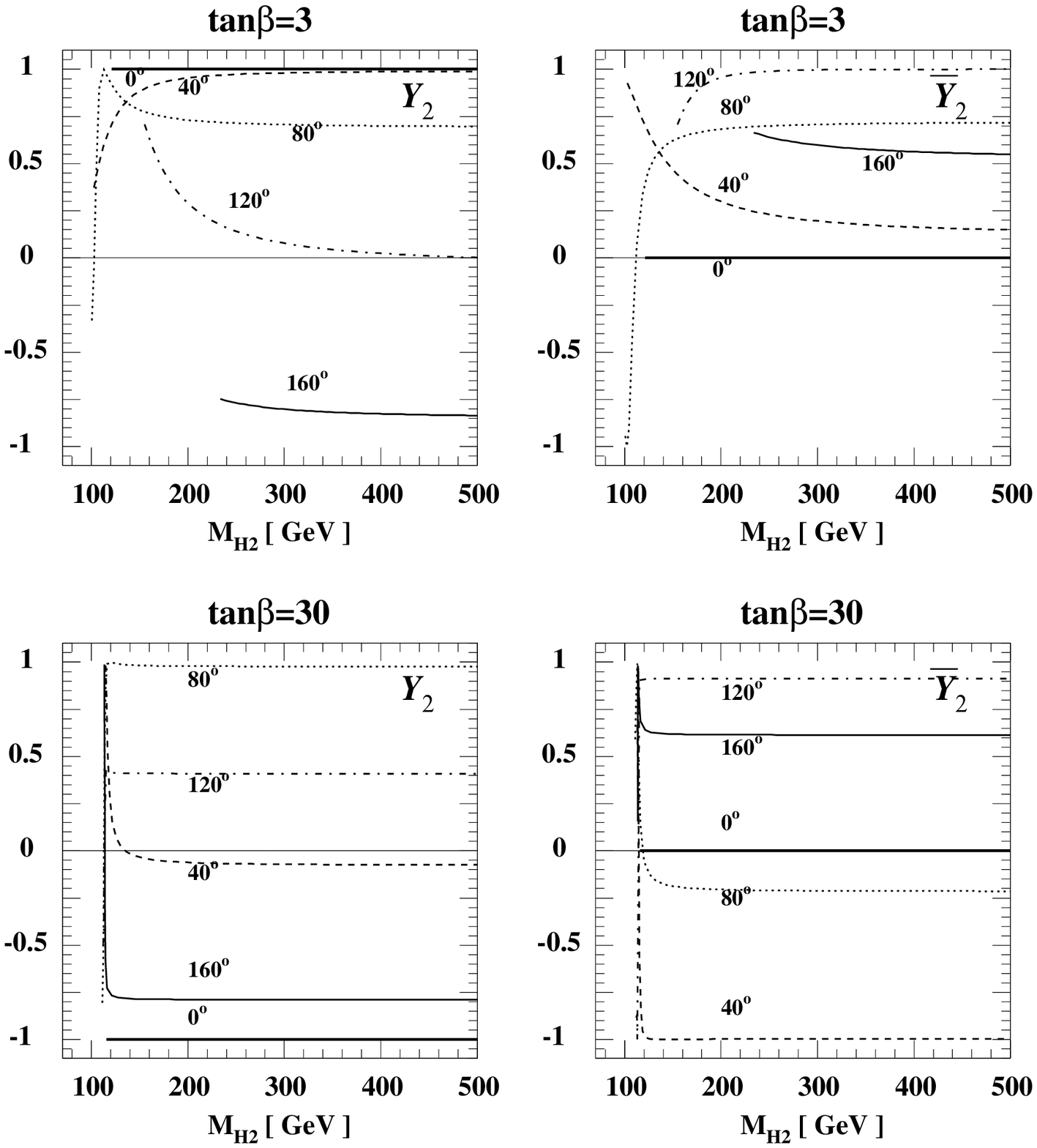,width=16cm,height=14cm}\hss}
 \end{center}
\caption{The CP--even and CP--odd TP asymmetries ${\cal Y}_2$ (left) 
          and $\overline{{\cal Y}_2}$ (right) versus $M_{H_2}$ with $\Phi=0^0, 
          40^0, 80^0, 120^0$, and $160^0$ for $\tan\beta=3$ (upper) and 
	  $\tan\beta=30$ (lower).} 
\label{fig4}
\end{figure}
\begin{figure}
 \begin{center}
 \hbox to\textwidth{\hss\epsfig{file=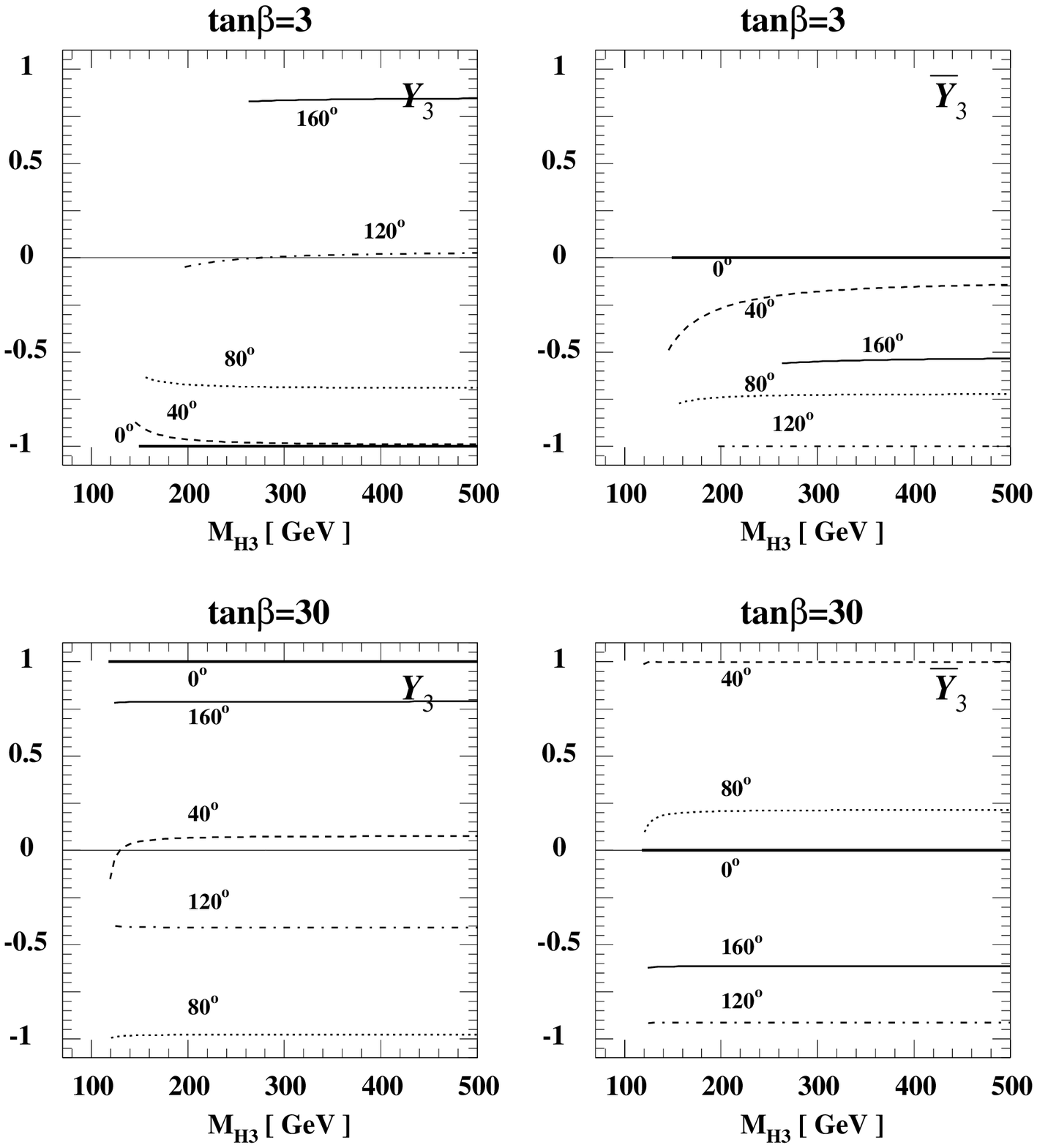,width=16cm,height=14cm}\hss}
 \end{center}
\caption{The CP--even and CP--odd TP asymmetries ${\cal Y}_3$ (left)
          and $\overline{{\cal Y}_3}$ (right) versus $M_{H_3}$ with 
          $\Phi=0^0, 40^0, 80^0, 120^0$, and $160^0$ for $\tan\beta=3$ (upper) 
	  and $\tan\beta=30$ (lower).} 
\label{fig5}
\end{figure}

\vfil\eject

\end{document}